# Improved Dynamic Response in Grid-Forming Converters with Current Limiting Control during Fault Conditions

Somayeh Mehri Boroojeni[a], Ehsan Sharafoddin[b],

[a] Department of Electrical Engineering, New Mexico State University, New Mexico, USA. Email: smehri@nmsu.edu

[b] Chief Executive Officer (CEO), Ertebat Andishan Sanaat Safahan Company, Isfahan, Iran. Email: ehsansharafodin@gmail.com

**Highlights**

- In this article, grid-forming converters are exposed to grid faults, and an error-mode mechanism is proposed for conserving the voltage-mode properties of grid-forming converters.
- By preserving the voltage-mode properties of grid-forming converters, converter currents are regulated and controlled.
- A dynamic virtual damping algorithm is suggested in this article to improve fault isolation.
- A dynamic virtual damping controller and fault mode for the grid-forming converter are modeled and simulated under very weak grid conditions.

**Abstract**

Modern power systems increasingly demand converter-driven generation systems that integrate seamlessly with grid infrastructure. Grid-based converters are particularly advantageous, as they operate in harmony with conventional synchronous machines. However, most existing research focuses on managing grid-forming converters (GFM) under normal conditions, often neglecting the converters' behavior during faults and their short-circuit capabilities. This paper addresses these gaps by introducing a power matching-based current limitation scheme, which ensures GFM converter synchronization while preventing over currents. It also highlights the limitations of grid-following techniques, which need to maintain robust grid-forming properties during fault conditions. Unlike conventional methods, no assumptions are made regarding outer power loops or droop mechanisms, and current references are immediately restricted to prevent wind-ups. A dynamic virtual damping algorithm is proposed to improve fault isolation further. This technique enhances fault-ride-through capability and maintains grid-forming properties even in weak grid conditions. The dynamic virtual damping controller and fault mode for GFMs are modeled and validated using detailed simulations in MATLAB. These results demonstrate that altering outer power sources, rather than internal structures, improves converter performance during faults, ensuring grid stability and reliability.

## 1. Introduction

The increasing integration of variable renewable energy sources, such as wind and solar, along with the rise of high-voltage direct current (HVDC) networks, is significantly transforming modern power systems [1]. This shift has led to a substantial increase in the use of power electronic converters, which now play a significant role in electricity generation. In these modern grids, power converters generate a large portion of electricity, while synchronous machines provide the remainder. However, grid control modes present several limitations—they cannot operate independently and negatively impact system response, raising concerns about the stability of converter-controlled power systems [2].

Grid-forming control, a significant breakthrough in power systems, holds the promise of overcoming these limitations. By empowering converters to function as voltage sources, grid-forming converters (GFMs) can mimic key features of synchronous generators, such as voltage and frequency control. However, GFMs face a critical challenge with overcurrent protection due to their voltage-source behavior, making them more susceptible to short-circuit faults than traditional synchronous machines [3], [4].

As GFMs are increasingly being considered as alternatives to synchronous generators (SGs) in power systems, the demand for reliable synchronization and current-limiting functionalities becomes paramount. These features are not just important, but crucial for maintaining GFM stability, particularly under weak grid conditions [5].

The transition to converter-based power systems involves systematically replacing large synchronous machines, which have traditionally provided critical stability and backup power during faults. These machines produce large short-circuit currents (6-8pu) during grid faults, while power electronics-based generators are limited by their current ratings, shifting the focus toward overall system efficiency [6]. In grid-connected systems, converters typically operate in grid-following mode, using phase-locked loops (PLLs), such as synchronous reference frame PLLs (SRF-PLLs), for synchronization [7]. However, under weak grid conditions or during faults, these methods can introduce low-frequency fluctuations and synchronization issues [8].

While grid-following converters are effective under stable conditions, they lack the robustness required during disturbances. This limitation has made GFMs an attractive solution for simulating the operation of synchronous machines and providing better support during grid disturbances [6]. However, power electronics-based systems lack the inherent rotational inertia that traditionally offers high stability and reliability during transients in synchronous machines [9], [10]. It makes the control strategies for converter-based systems crucial for maintaining stability during grid disturbances.

Several control strategies for GFMs have been developed, including droop control [11], [12], virtual synchronous machines (VSM) [13], virtual synchronous generators [14], power synchronization control (PSC) [15], and synchronous power controllers (SPC) [16]. These methods regulate voltage and frequency at local connection points, ensuring system stability. However, while GFMs are well-studied under normal conditions, their transient response and behavior during faults remain less understood, primarily due to the challenges of replicating the short-circuit characteristics of synchronous machines with current-limited semiconductor devices [17], [18].

Various methods have been proposed to mitigate over currents, yet each presents some challenges. For example, Oureilidis et al. [19] suggested a method that switches from grid-forming to grid-following mode, enabling the use of grid-following converters' current limitation capabilities. However, this approach results in a sudden shift in phase angles, reducing robustness and introducing additional control complexity. Virtual impedance-based strategies, such as those proposed by Lin et al. [20], limit the voltage reference to minimize the output current. However, the effectiveness of virtual impedance depends on precise grid impedance calculation, which is often difficult to determine. Liu et al. [21] introduced a dynamic virtual impedance method, but identifying fault locations and line impedance remains a challenge, making current limitations difficult.

Voltage limitation strategies have also been explored. Chen et al. [22] proposed adjusting the GFM converter's output voltage to match the point of standard coupling (PCC), but this method struggles with frequency changes during overloads. Similarly, Huang et al. [23] introduced a power angle-based adaptive overcurrent mitigation approach that does not require a phase-locked loop (PLL). However, this method faces reliability issues under unstable grid conditions.

In the event of grid disturbances, GFMs must remain operational so that, once the fault is cleared, they can resume normal functioning. Our study underscores the practical and economical need for a current-limiting method that preserves the GFM control structure while limiting current during faults. It examines how outer control loops affect converter performance and proposes enhanced current-limiting control schemes for GFMs during fault conditions. The proposed solution limits converter current references and adjusts output power references to prevent wind-ups in outer loops, thereby preserving the dynamics of GFMs during faults.

Furthermore, the current paper presents a power matching-based current limitation scheme that ensures GFM synchronization during faults while preventing over currents. Unlike prior approaches, our solution makes no assumptions about outer power loops or droop mechanisms and focuses on avoiding wind-ups while maintaining stability. Additionally, a dynamic virtual damping algorithm is proposed to enhance fault isolation and ensure the GFM retains its grid-forming properties during disturbances.

## 2. The proposed grid-forming converter method

Fig. 1 shows a voltage source converters (VSC) system with a grid-tied and grid-formed model connected to the outside Thevenin system. Table 1 summarizes the significant elements of the grid-forming converter illustrated in Fig. 1. Fig. 2 presents an in-depth description of the grid-forming control architecture called SPC.

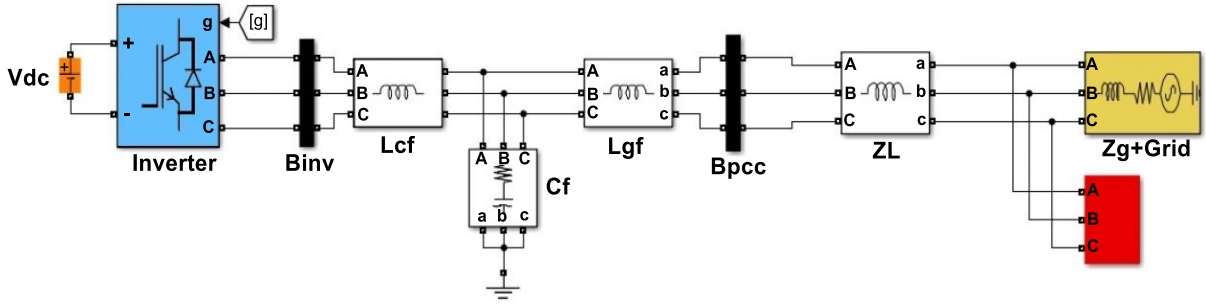

**Fig**. 1. Grid-tied grid-forming VSC coupled to an outside Thevenin model grid with a symmetrical fault near the variable-frequency ($v_f$) bus.

Grid-forming converters simulate the physical and electrical properties of conventional synchronous generators. The physical part of the synchronous generator helps stabilize the network frequency by simulating torque and resonance. Modeling the electrical element based on simulated conductor resistance enables the distribution and transmission of power with the network.

**Table 1**

The primary parameters of the system in Fig. 1.

| Symbol | Description | Value |
| --- | --- | --- |
| $V_{dc}$ | dc-Link Voltage | 400 V |
| $f_0$ | Rated Frequency | 50 Hz |
| $f_{sw}$ | Switching frequency | 10 kHz |
| $f_s$ | Sampling frequency | 10 kHz |
| $L_{cf}$ | Convertor-side inductor | 0.05pu |
| $L_{gf}$ | Grid-side inductor | 0.06pu |
| $C_f$ | Filter Capacitor | 0.02pu |
| $Z_L$ | Line impedance | 0.02j-0.5j pu |

In Fig. 2, the active and reactive power values produced by the *P&Q Calculation* block are measured using currents passing through the inductor and applied to the *PQ Controller* block. This block, the *PQ Controller*, calculates active and reactive power based on Fig. 1 and *PI* regulators. It receives the network's nominal frequency values and nominal phase voltage range.

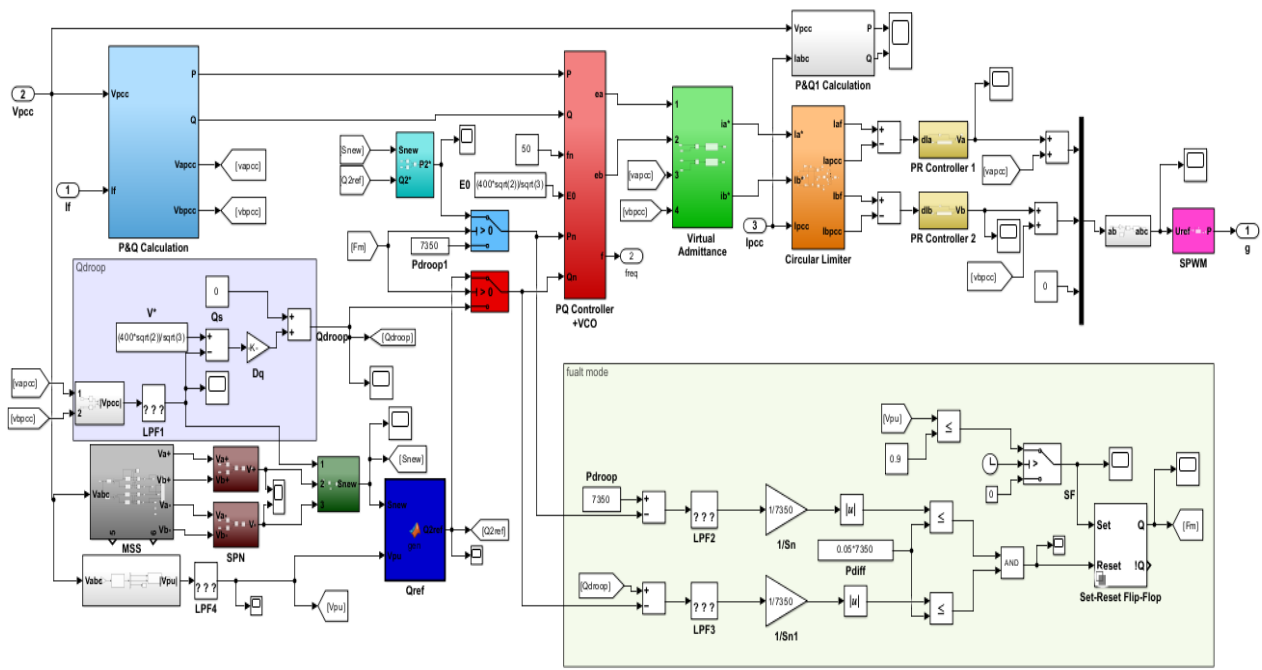

**Fig. 2**. The SPC control scheme block diagram. Frequencies in PLCs are modified in response to changes in active power from droop controllers. It has a specified natural oscillating frequency and an adjustment factor.

Trial and error are used in this part to modify the proportional-integral parameters of the controller. From this block, both the network frequency and the virtual EMF value are obtained. Next, the EMF values and connection point voltages are applied to the Virtual Admittance block, calculated in Eq. (3). These currents are then calculated in a Circular Limiter block using Eq. (13). In the next step, the results of both current comparisons are applied to PR controller blocks. These blocks are responsible for setting connection point currents and producing sub-voltages. These voltages are collected from the common connection point and produce reference voltage signals. The reference voltage signals are first transferred from the stationary device to the *abc* device. They are then sent to the SPWM block, which generates the activation signals of the inverter keys based on the SPWM rule. Finally, the generated signals are transferred to the inverter keys.

Eqs. (1) and (2), as well as the fault mode schemes in Fig. 2, are used to generate active and reactive power references based on droop or from power reference modes. Positive and negative sequence voltages are obtained from the MSS block. The SPN block then calculates the voltage range and signal of the common connection point. In the case of droop mode, reactive power is calculated based on Eq. (2). The Power Loop Controllers (PLCs) play a crucial role in providing the converter with the technical characteristics of simulated machines and perform calculations relating power harmony to approximate rotation frequency. In the process of modeling machines, resisting power is controlled via a resisting power controller (RPC).

A virtual machine with simulated rotational frequency and voltage magnitude can calculate the inner electromagnetic energy. As a result of the deviations in the network voltages and frequencies, two outer loops calculate a reference for passive power. Similarly to the generator and automatic voltage controller of a synchronous machine, drop converters are theoretically described as follows:

$$P^* = Ps + (\omega^* - \omega)D_P \tag{1}$$

$$Q^* = Qs + (V^* - |V_{PCC}|)D_Q \tag{2}$$

where Ps and Qs are the outer control parameters for active and reactive power, $\omega$ is the simulated resonance frequency provided by the PLC, and DP and DQ are drop increases for active and reactive power. Fig. 2 illustrates the virtual admittance block, which determines its rotor coils to maintain an inductive output. It allows precise and uncoupled control of active and reactive power. As a result, the simulated or virtual impedance scheme is formed as follows:

$$i^*_{\alpha\beta} = \frac{e_{\alpha\beta} - v_{\alpha\beta}}{R_v + sL_v} \tag{3}$$

$R_v$ and $L_v$ are the simulated power supply and generator resistances, respectively; $e$ is the modeled EMF derived from two external power circuits, and $v$ is the voltage obtained at the point of common coupling (PCC). Simulated resistance should be set substantially lower than the frequency of the inner current controller since it operates in a parallel loop with a virtual resistance. According to Shi et al. [23], the calculated impedance should equal 0.3pu of the converter's maximum impedance. As an example, this is the typical response value for a grid-powered synchronous machine. In this way, the simulation resistance is determined to obtain the simulated inductance circuit's threshold frequency. The internal current controller can be used to simulate an impedance of 0.1pu by selecting a threshold frequency ten times lower than the internal current controller.

When the virtual admittance of the virtual system is selected in a way that the capacitive impedance seen at the converter inputs is high, it paves the way for the calculation of three-phase power transmission. This transmission, which occurs between the receiving grid and the virtual system, is a vital part of the system's operation and performance.

$$P = \frac{3}{2}\frac{EV_g \sin(\delta)}{X} \approx \frac{3}{2}\frac{EV_g}{X}P_{max}\delta \tag{4}$$

$$Q = \frac{3}{2}\left(\frac{E^2 - EV_g \cos(\delta)}{X}\right) \approx \frac{3}{2}\frac{E(E - V_g)}{X} \tag{5}$$

In Eqs. (4) and (5), the phase angle difference between two data points is calculated as the value of the phase angle difference. It also calculates $X$ as the overall output impedance between two data points, Pmax as the nominal active power of three phases, and $E$ as the maximum voltages at the sending and receiving ends. An SPC is similar to a synchronous machine in that it relies on the internal synchronization mechanism of AC networks. The power-driven synchronization mechanism in PLCs provides the synchronization angle/frequency by regulating the active power error. Simple methods are typically used to simulate synchronous machine properties using the alternating equation:

$$\frac{2HS_n}{\omega_0}\frac{d^2\delta}{dt} = P_m - P_e - D\omega_0 \frac{d\delta}{dt} \tag{6}$$

There are six parameters in Eq. (6): $H$ is the torque constant, $S$ is the output power, $D$ is the total damping amount, $P_m$ is the mechanical power, $P_e$ is the electrical power, $\omega_0$ is the electrical angular frequency, and $\delta$ is the load direction of the machine in connection with the grid. A *PI* controller is used in the PLC instead of the swing equation. When there is a deviation in frequency, the damping amount of the swing equation produces a steady-state droop effect [6]. Fig. 3 shows PLC dynamics as follows:

$$\frac{P(s)}{P^*(s)} = \frac{P_{max}K_{pp}s + K_{ip}P_{max}}{s^2 + K_{pp}P_{max}s + K_{ip}P_{max}} = \frac{2\zeta\omega_N s + \omega_N^2}{s^2 + 2\zeta\omega_N s + \omega_N^2} \tag{7}$$

In order to reproduce the inertia constant and damping proportion of second-order reactions, the controller gains should be chosen as follows:

$$K_{ip} = \frac{\omega 0}{2HS_n}, K_{pp} = \zeta\sqrt{\frac{2\omega_0}{HS_n P_{max}}} \tag{8}$$

The torque constant is regulated at $H \approx 2\text{–}5$ s.

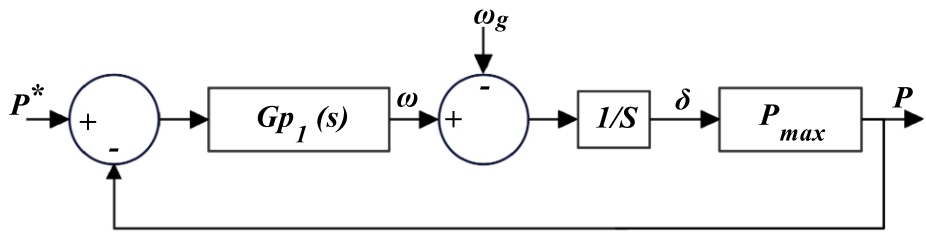

**Fig. 3.** In this diagram, a *PLC* block with a *PI* controller is shown, which determines the correlation between a power fault and the virtual frequency as a result of an active power fault.

## 3. Control challenges with grid-forming converters under grid faults

The present study defines strategies for current limits, resulting in a recommended technique for grid-forming control of converters that has been shown to significantly reduce converter current without compromising the standard method of grid-forming. Our proposed grid-forming converter strategy described in the previous section, must be combined to achieve this result: internal current limitation and external power reference adjustment. Furthermore, once the fault has been removed, the grid-forming converter can be improved using a dynamic simulating damping controller.

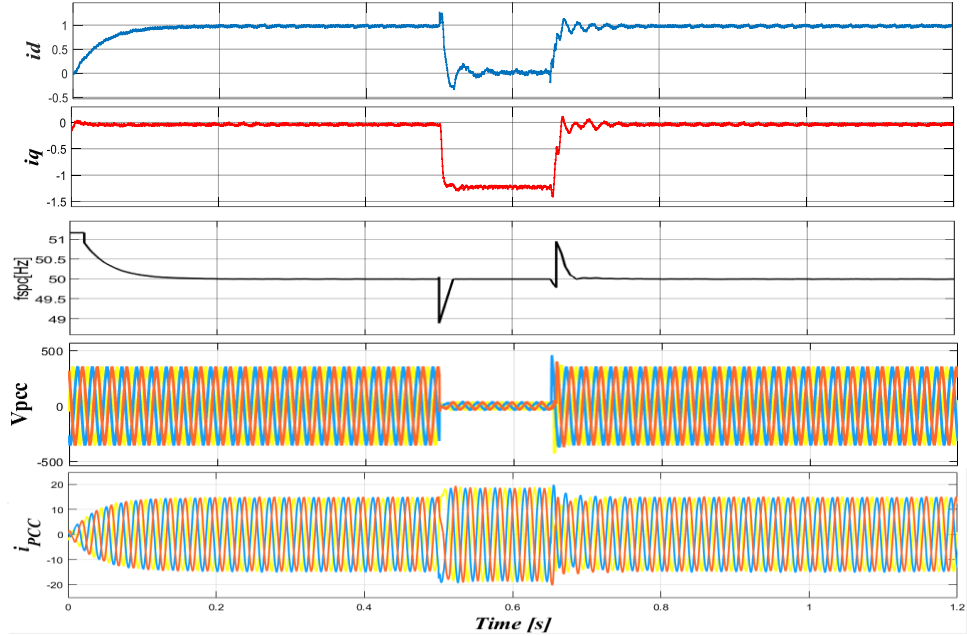

**Fig. 4**. A simulation of SPC fault response under a symmetrical fault with magnitudes of 0.3pu and impedances of 0.04pu. Actual and reference values for the dq-axes currents are shown here in blue and red, respectively.

**Table 2**
SPC structure parameter in Fig. 2.

| Grid-Forming Controller (SPC) | Parameters |
|---|---|
| Droop control | $D_p = 0$, $D_q = 108$ Var/V |
| Virtual Admittance | $R_v = 0.107$ pu, $L_v = 0.375$ pu |
| Current Controller | $K_p = 25$, $K_r = 2000$ |
| PLC | $H = 2$, $\zeta = 0.5627$, $K_{pp} = 1.16\ e^{-3}$, $K_{ip} = 5.86\ e^{-3}$ |
| RPC | $\zeta = 0.5627$, $\omega_N = 25.45$ rad/s, $L_{eq} = 29.5$ mH, $E_{en} = 400$ v |

Table 2 presents the controller parameters for a grid-forming structure during a symmetrical three-phase fault, as simulated in Fig. 4, which highlights the SPC's control challenges. In this scenario, the

Thevenin grid model applies three-phase voltages that are directly controlled during a fault event. Fig. 4 illustrates that the grid-forming structure inherently supports the PCC voltage without modifications during a fault. This voltage support is primarily attributed to a reactive current injection of 7.14pu. During fault events triggered by droop controllers, outer power references are disregarded. Fig. 4 demonstrates the system's response to a balanced fault using the proposed control technique. As the fault is cleared, the actual active power deviates from the expected value, leading to a frequency response governed by the simulated mechanical properties of the virtual machine.

During this brief surge in virtual resonating frequency, the converter adjusts its load to maintain synchronization with the external grid. When this control method is applied, the current injected during the fault complies with grid code requirements. It allows post-fault performance to be significantly improved without exceeding current limits and without relying on PLL-based grid-following methods during faults. As Mukherjee et al. suggested [24], power distribution control for converter current shifts to current mode control at the onset of a grid fault. However, an alternative PLL is necessary to ensure the converter remains connected to the grid under fault conditions. Ma et al.'s research [25] highlights the critical role of the PLL in converter accuracy, especially under low-voltage conditions. Therefore, the method proposed in Fig. 2 is advantageous, as it limits converter currents while adhering to grid code requirements, all without altering the fundamental control scheme.

Despite being a voltage-regulated structure, the SPC only requires a current reference to supply the voltage necessary to meet PLL and RPC demands. Unlike grid-following converters, SPCs function as voltage sources regulated by their magnitude and frequency. Under short-circuit fault conditions, high current values are injected to regulate the voltage source and stabilize the voltage level. Protecting semiconductor devices from overcurrent when using voltage source converters (VSCs) is crucial. VSCs must limit the current since their current reference can exceed the safe operating limits. The SPC can support the grid if there is a power reserve, which improves the network's transient stability. However, grid faults may compromise the converter's synchronization stability [26]. Given the limited overcurrent capabilities of GFM converter power devices, they must restrict their terminal current during symmetrical faults [27]. In such cases, outer control loops may fail to meet voltage reference requests due to the low current amplitude. The effectiveness of voltage mode control structures during symmetrical grid faults is typically assessed by evaluating different current-limiting techniques.

Current limitation becomes essential during grid faults when the converter risks overloading. One approach is to oversize the converter to accommodate higher currents; however, this increases the overall cost, which is undesirable. Alternatively, the current limitation can be achieved by reducing virtual admittance. However, two critical issues must be addressed before implementing this approach. First, the maximum current admittance value depends on voltage drop and the control response under varying fault conditions, necessitating adjustments to the admittance value to reach a specified maximum current. Reducing the admittance until the maximum current is achieved can be an effective strategy, but the converter's current response must be consistent across fault conditions.

Implementing a direct current limiter is a more straightforward and intuitive approach when using reference values. In another scenario, if the inductance of the virtual admittance structure increases during a fault, the current references may experience a bias due to the inability to change the inductor current instantly. This results in an undesirable decaying DC component in the AC signal at the current connection point. If the sole objective during a fault is to limit converter currents, adjusting the current reference remains the most effective method. Introducing a high-power resistor would reduce the peak signal and output current in a control system that operates with periodic signals. Alternatively, the fixed current reference can be calculated by employing a circular resistor as follows:

$$i_{\alpha\beta}^* = \begin{cases} i_{\alpha\beta} = \dfrac{I_{\lim}}{\sqrt{i_\alpha^2 + i_\alpha^2}} & if \ \sqrt{i_\alpha^2 + i_\alpha^2} > I_{\lim} \\ i_{\alpha\beta} & \text{otherwise} \end{cases} \qquad (9)$$

where $I_{lim}$ represents 1.2pu of the standard converter current. By generating additional power, the external control processes intensify the voltage, while the PLC limits its output power to 1.2 watts.

A few other things should be considered as well. Despite injecting capacitive reactance current, the converter does not synchronize with grid code because current discharged during fault periods decreases voltage support. The outer loop post-fault response is extended and unsatisfactory due to high-power references and saturation of current references. During this stage, the active power is gradually raised to 1.2pu until it equals the current limiter power limits. This results in the converter returning to the steady-state condition. Even with converter currents restricted in this study, fault detection is required, particularly after a fault has occurred. With the method described, it is necessary to modify the outer power loops to account for the converter's low tolerance for input currents, resulting in saturation. As a solution to this problem, it should be possible to change the active and passive power references following a fault to control the converter output current [28]. As a result, the permissible converter power can be modified when a fault appears as follows:

$$S_{new} = \sqrt{\frac{3}{2}} \cdot \frac{V^+ - V^-}{V_b} S_n \tag{10}$$

In Eq. (10), $V+ = (v\alpha+2 + v\beta+2)1/2$ and $V- = (v\alpha-2 + v\beta-2)1/2$ represent the strength of the positive and negative order voltages, respectively. With $V- = 0$, a balanced fault is calculated as follows:

$$S_{new} = V_{pu} S_n \tag{11}$$

Using grid codes, we can calculate the inertia power reference as:

$$Q^* = \begin{cases} \text{Voltage Droop if } V_{pu} > 0.9 \\ 2S_{new}(1 - V_{pu}) \text{ if } 0.5 < V_{pu} < 0.9 \\ S_{new} \quad \text{otherwise.} \end{cases} \tag{12}$$

By combining the adjusted noticeable power with the response power reference, the active power reference that prevents destroying excessive currents is calculated as follows:

$$P^* = \sqrt{S_{new}^2 - Q^{*2}} \tag{13}$$

When the converter's output power exceeds the active power reference, the active power reference is zeroed. In contrast, the reactive power reference equals $S_{new}$. The power limiting technique mentioned in Eq. (16) is tested during a balanced fault. As a result, the maximum injection current values are reduced to roughly 3pu and firmly decreased to 1pu as the power reference dictates. Despite that, over-currents persist for more than ten initial periods because of the RPC's and PLC's delayed reaction time.

## 4. Results and simulations

The current limitation approach proposed here is derived from data collection and methods described here that are specific to the current limitations. Restricting the converter current during a fault protects the central part of the grid-forming control. The inputs of current limitations integrated determine the output power reference modification. Furthermore, an enhanced fault restoration technique for the grid-forming converter is proposed using a dynamic virtual damping controller. Matrix Laboratory (MATLAB) is used to simulate the results of this method.

*4.1. Current limitation and reference power*

It is suggested the reference power be modified by the voltage dip during the fault (Eq. (16)) and circular current limitations (Eq. (13)). Consequently, the external power regulators' parameters are modified to meet the grid code criteria, thereby controlling the current. Therefore, the assistance and the suggested fault-mode converter include reformulating and applying the virtual limiter to produce accurate, current limits, as well as realizing that the power adjustment method from [28] is suitable for grid-forming converters with external power loops to prevent noise and overloading. In addition, the results of the SPC's drop test and the changes between the two power references are also discussed. Fig. 2 illustrates the suggested scheme, where FM is the fault-mode signal that controls the power references between the outer drop controllers and the power reference according to grid code criteria. An abnormal fault signal (SF) is emitted when the fixed frame voltage value exceeds 0.9pu. In this case, FM shifts to fault-mode control when the reaction time drops to 1 millisecond.

As long as fault-mode control is preserved, the approach in Fig. 2 will be applied until the offset between the controller's active and passive power references and the fault-mode control is preserved. When fault-mode control has been recovered, the $P_{diff}$ value will fall below it in Eqs. (12) and (13).

Consequently, fault-mode control is enabled after fault recovery based on $P_{diff}$, droop coefficients, and failure detection from external networks. Fault-mode control is applied 150*ms* after the fault is resolved, assuming general droop factors and $P_{diff}$ equals 5%. Due to the low transmission rates of the external PLC and RPC and the absence of fault-mode control, the distinction between the two controllers is insignificant without fault-mode control. As a result of the introduced currents, the switchback of the drop controllers occurs automatically without interfering with the system's function. Because of this configuration, the properties of the simulated model persist under regular and fault scenarios, as well as limiting the converter current.

*4.2. Enhanced fault recovery using dynamic damping*

Fig. 4 shows an inevitable fluctuation in the level of dampness during restoration. As the converter voltage increases, the SCR tends to react less to these fluctuations since the input currents from the converter are more susceptible to the PCC voltage. Dynamic damping techniques and fault control can accelerate the recovery cycle. The essential strategy is temporarily reducing the conductivity in the simulated permeability model over the recovery process, as shown in Fig. 5.

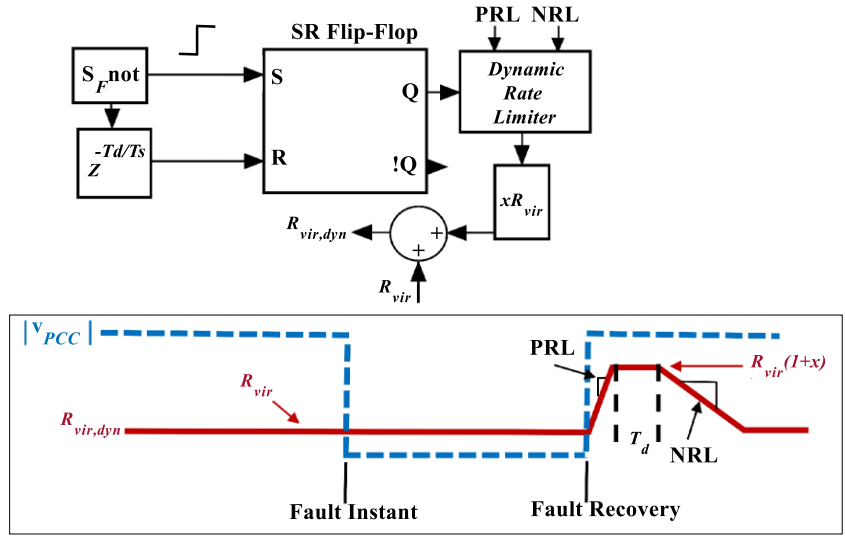

**Fig. 5**. The proposed dynamic damping scheme. During fault recovery, the virtual damping is temporarily increased to provide additional damping. Based on the PRL, the dynamic virtual impedance, $R_{vir,dyn}$, is immediately raised to $R_{vir}(1+x)$ following voltage restoration. According to the acceleration rate determined by the NRL, the virtual impedance will be lowered to $R_{vir}$ once the delay of $T_d$ has been applied.

In response to the fault removal, the SR flip-flop output reaches its maximum, driving the simulated impedance to jump from $R_{vir}$ to $R_{vir}(1+x)$ at the rate determined by the positive rate limiter (PRL). During the first Td seconds, enhanced damping is preserved, and the impedance is reduced to its post-fault condition at a negative rate limiter (NRL).

Fig. 5 illustrates the PRL ramp at a low angle to make it more visible. In other words, when PRL is set to a slope of *10000*, the simulated impedance will increase in one sampling period. As a result, virtual admissions start at $R_{vir}(1+x)$ and ramp down to $R_{vir}$ in 10*ms* at the NRL. It is possible to calibrate x to achieve suitable fault recovery performance manually. In this case, simulated damping must be applied, as shown in Fig. 5.

Fig. 6 illustrates how the dynamic damping controller in Fig. 5 caused oscillations in the dq-axis currents to be significantly reduced during the recovery process by applying the dynamic damping controller to Fig. 6.

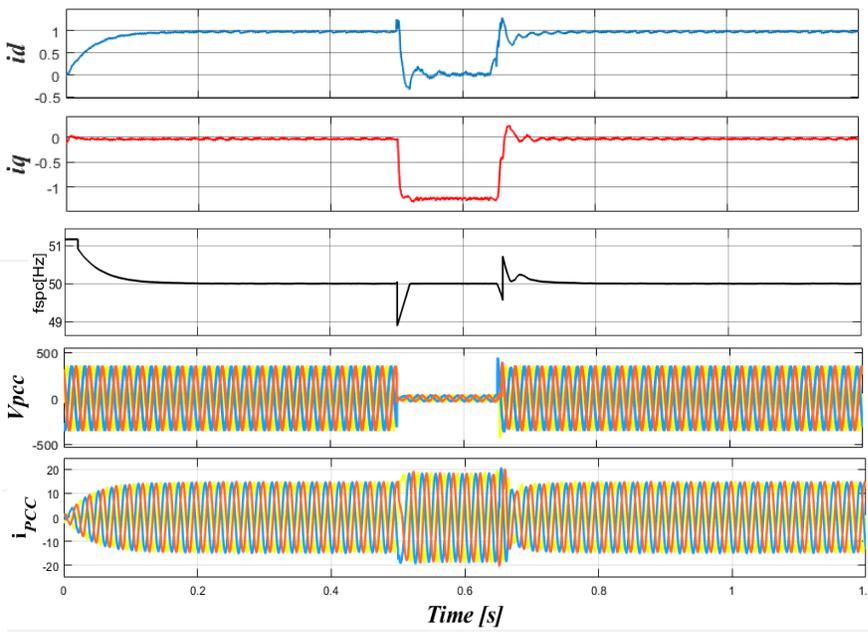

**Fig. 6**. A simulation model analysis identical to Fig. 4 but with dynamic damping applied during fault recovery instead of static damping, as shown in Fig. 5, where x = 1. A comparison of the actual (blue) and reference (red) currents is shown on the dq-axes.

Fig. 7 shows the fault recovery methods utilizing different dynamic damping values. After recovery, the dq-axis currents are shown at 0*s* after clearing the fault.

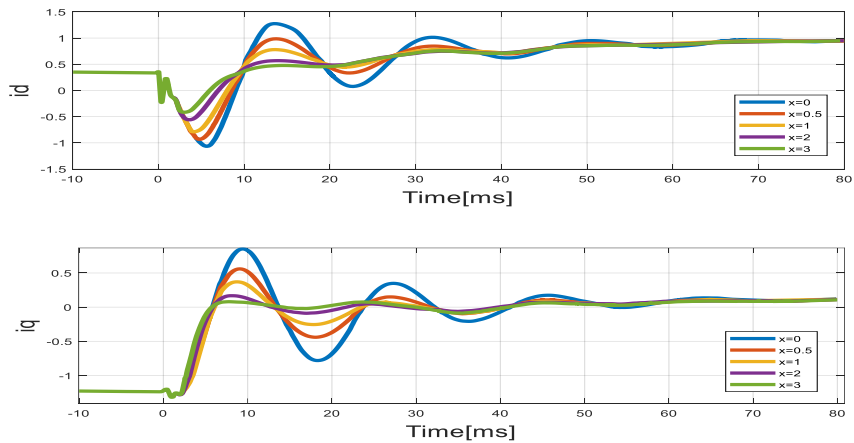

**Fig 7**. As shown in Fig. 4, dynamic damping affects *idq* during fault recovery starting at 0 s. Fig. 5 illustrated how x affects dynamic damping.

By increasing imaginary resistance, dynamic damping controls can be significantly reduced in both upshoot and downshoot in $i_d$ and $i_q$. In this case, the overshoot in $i_q$ can be eliminated. By controlling $i_q$ properly, voltage recovery on PCCs can be enhanced since a positive $i_q$ will cause the converter to behave as an inductor, enhancing voltage recovery.

## 4.3. Weak-Grid Performance

Compared to Shi et al.'s study [23], our synchronous power controller describes a detailed and practical fault-mode controller for grid-forming converters during failure situations based on a fault-mode controller and dynamic dampening structure. Shifting from grid-following PLL-synchronized to grid-following PLL-synchronized regulates converter currents during faults. In the event of a fault, the grid-forming controller is switched back to a feedback-tracking controller, which then aligns the phase angles of the two controllers to ensure seamless switching.

Fig. 8 shows the output produced by the model-based scheme in the last simulation, where SCR = 2. Despite the nominal power injection, the high line impedance requires a significant amount of reactive current. When a grid fault occurs, the injected current is above average, resulting in complicated conditions.

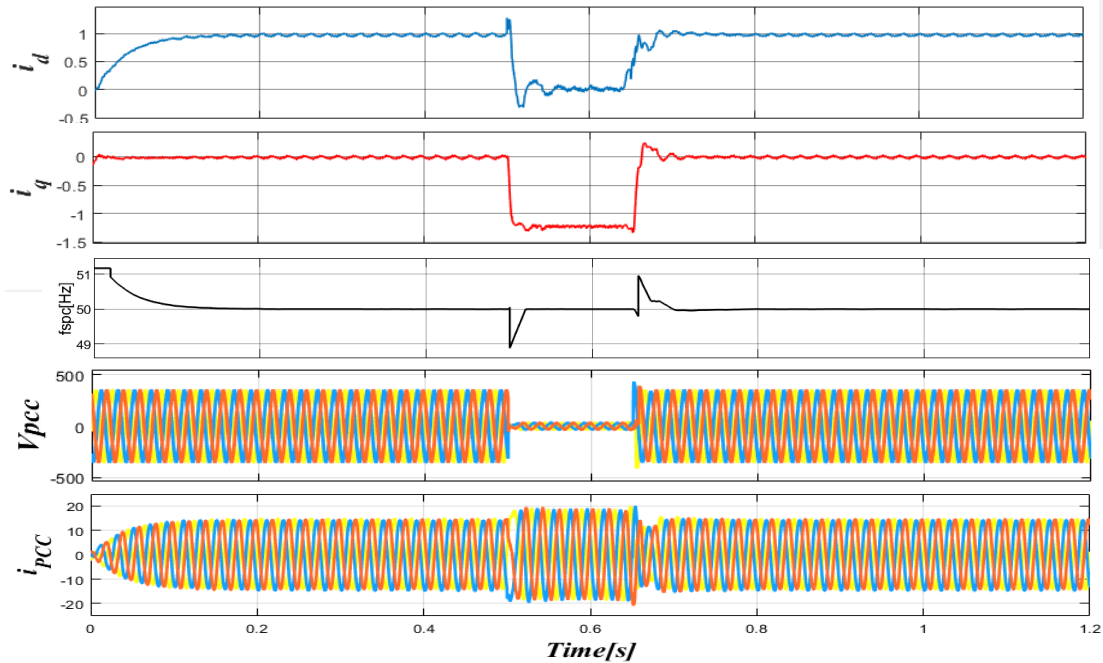

**Fig. 8**. The simulation of the proposed SPC using dynamic damping with x = 2 during a symmetrical fault with a voltage magnitude of 0.3pu. In Fig. 2, the reference power calculation was shown using the control structure. For the dq-axes currents, the actual (blue) and the reference values (red) are visualized. (a) SCR = 5. (b) SRC = 2.

The dq-axes currents reach their reference values quickly in a stable process, and the recovery mechanism occurs with low oscillations and barely noticeable overshoots in the q-axes current. In the axes current, a sudden increase in voltage and currents can be noted around 20ms after the fault occurs due to the abnormal and temporarily unbalanced currents and voltages. Therefore, the developed structure effectively controls converter currents and transitions between the proposed fault-mode controller and the proposed fault-mode converter, even during quite unstable grid conditions.

## 5. Conclusion

Due to their grid-supportive characteristics, grid-forming converters are expected to become highly desirable in next-generation power systems dominated by power electronics. However, unlike synchronous generators, these converters rely on current-sensitive semiconductor devices, limiting their ability to maintain synchronous generator-like behavior during grid faults. This paper has outlined the critical restrictions of current grid-forming converters and proposed methods to address them. In addition to offering significant advantages, the proposed solution adheres to grid code requirements, eliminating the need for fundamental modifications to the control structure during fault conditions. In order to enhance fault isolation, a dynamic virtual damping algorithm has been introduced, ensuring that grid-forming converters can continue to support grid functions even under normal operating conditions. By adapting the external power control rather than altering the internal structure, the system's fault-ride-through capability is significantly improved. The dynamic virtual damping controller and fault mode for grid-forming converters have been modeled and simulated under weak grid conditions, demonstrating enhanced stability and performance during grid disturbances. This research is crucial for advancing the reliability and resilience of future power systems, particularly as the integration of renewable energy sources and power electronics increases. The findings can guide the development of more robust and efficient grid-forming converters, contributing to the overall stability and sustainability of modern electric grids.

*Notes*

The authors did not receive support from any organization for the submitted work.

**Financial interests**: The authors declare they have no financial interests.

*Data availability statement*

Data will be made available in a repository on acceptance.

# Reference

[1] Zarei SF, Mokhtari H, Ghasemi MA, Blaabjerg F. Reinforcing fault ride through capability of grid forming voltage source converters using an enhanced voltage control scheme. IEEE Transactions on Power Delivery. 2018 Jun 4;34(5):1827-42. https://doi.org/10.1109/TPWRD.2018.2844082

[2] Çelik D, Meral ME. A flexible control strategy with overcurrent limitation in distributed generation systems. International Journal of Electrical Power & Energy Systems. 2019 Jan 1;104:456-71. https://doi.org/10.1016/j.ijepes.2018.06.048

[3] Çelik D, Meral ME. A coordinated virtual impedance control scheme for three phase four leg inverters of electric vehicle to grid (V2G). Energy. 2022 May 1;246:123354. https://doi.org/10.1016/j.energy.2022.123354

[4] Qoria T, Wang X, Kadri R. Grid-forming control VSC-based including current limitation and re-synchronization functions to deal with symmetrical and asymmetrical faults. Electric Power Systems Research. 2023 Oct 1;223:109647. https://doi.org/10.1016/j.epsr.2023.109647

[5] Liu T, Wang X, Liu F, Xin K, Liu Y. A current limiting method for single-loop voltage-magnitude controlled grid-forming converters during symmetrical faults. IEEE Transactions on Power Electronics. 2021 Oct 27;37(4):4751-63. https://doi.org/10.1109/TPEL.2021.3122744

[6] Taul MG, Wang X, Davari P, Blaabjerg F. Current limiting control with enhanced dynamics of grid-forming converters during fault conditions. IEEE Journal of Emerging and Selected Topics in Power Electronics. 2019 Jul 29;8(2):1062-73. https://doi.org/10.1109/JESTPE.2019.2931477

[7] Fu Q, Du WJ, Su GY, Wang HF. Dynamic interactions between VSC-HVDC and power system with electromechanical oscillation modes-a comparison between the power synchronization control and vector control. In 2016 IEEE PES Asia-Pacific Power and Energy Engineering Conference (APPEEC) 2016 Oct 25 (pp. 949-956). IEEE. https://doi.org/10.1109/APPEEC.2016.7779635

[8] Taul MG, Wang X, Davari P, Blaabjerg F. An overview of assessment methods for synchronization stability of grid-connected converters under severe symmetrical grid faults. IEEE Transactions on Power Electronics. 2019 Feb 1;34(10):9655-70. https://doi.org/10.1109/TPEL.2019.2892142

[9] Ulbig A, Borsche TS, Andersson G. Impact of low rotational inertia on power system stability and operation. IFAC Proceedings Volumes. 2014 Jan 1;47(3):7290-7. https://doi.org/10.3182/20140824-6-ZA-1003.02615

[10] Zhong QC. Power-electronics-enabled autonomous power systems: Architecture and technical routes. IEEE Transactions on Industrial Electronics. 2017 Mar 2;64(7):5907-18. https://doi.org/10.1109/TIE.2017.2677339

[11] D'Arco S, Suul JA. Virtual synchronous machines—Classification of implementations and analysis of equivalence to droop controllers for microgrids. In 2013 IEEE Grenoble Conference 2013 Jun 16 (pp. 1-7). IEEE. https://doi.org/10.1109/PTC.2013.6652456

[12] Rocabert J, Luna A, Blaabjerg F, Rodriguez P. Control of power converters in AC microgrids. IEEE transactions on power electronics. 2012 May 15;27(11):4734-49. https://doi.org/10.1109/TPEL.2012.2199334

[13] Chen Y, Hesse R, Turschner D, Beck HP. Improving the grid power quality using virtual synchronous machines. In 2011 international conference on power engineering, energy and

[13]    electrical drives 2011 May 11 (pp. 1-6). IEEE. https://doi.org/10.1109/PowerEng.2011.6036498

[14]    Wu H, Ruan X, Yang D, Chen X, Zhao W, Lv Z, Zhong QC. Small-signal modeling and parameters design for virtual synchronous generators. IEEE Transactions on Industrial Electronics. 2016 Mar 16;63(7):4292-303. https://doi.org/10.1109/TIE.2016.2543181

[15]    Zhang L, Harnefors L, Nee HP. Interconnection of two very weak AC systems by VSC-HVDC links using power-synchronization control. IEEE transactions on power systems. 2010 May 24;26(1):344-55. https://doi.org/10.1109/TPWRS.2010.2047875

[16]    Zhang W, Remon D, Rodriguez P. Frequency support characteristics of grid-interactive power converters based on the synchronous power controller. IET Renewable Power Generation. 2017 Mar;11(4):470-9. https://doi.org/10.1049/iet-rpg.2016.0557

[17]    Guerrero JM, Matas J, de Vicuna LG, Castilla M, Miret J. Decentralized control for parallel operation of distributed generation inverters using resistive output impedance. IEEE Transactions on industrial electronics. 2007 Mar 19;54(2):994-1004. https://doi.org/10.1109/TIE.2007.892621

[18]    Shuai Z, Huang W, Shen C, Ge J, Shen ZJ. Characteristics and restraining method of fast transient inrush fault currents in synchronverters. IEEE Transactions on Industrial Electronics. 2017 Jan 16;64(9):7487-97. https://doi.org/10.1109/TIE.2017.2652362

[19]    Oureilidis KO, Demoulias CS. A fault clearing method in converter-dominated microgrids with conventional protection means. IEEE Transactions on Power Electronics. 2015 Sep 14;31(6):4628-40. http://dx.doi.org/10.1109/TPEL.2015.2476702

[20]    Lin X, Liang Z, Zheng Y, Lin Y, Kang Y. A current limiting strategy with parallel virtual impedance for three-phase three-leg inverter under asymmetrical short-circuit fault to improve the controllable capability of fault currents. IEEE Transactions on Power Electronics. 2018 Nov 1;34(8):8138-49. https://doi.org/10.1109/TPEL.2018.2879191

[21]    Chen J, Prystupczuk F, O'Donnell T. Use of voltage limits for current limitations in grid-forming converters. CSEE Journal of Power and Energy Systems. 2020 Feb 13;6(2):259-69. https://doi.org/10.17775/CSEEJPES.2019.02660

[22]    Narula A, Imgart P, Bongiorno M, Beza M, Svensson JR, Hasler JP. Voltage-Based Current Limitation Strategy to Preserve Grid-Forming Properties Under Severe Grid Disturbances. IEEE Open Journal of Power Electronics. 2023 Feb 20;4:176-88. https://doi.org/10.1109/OJPEL.2023.3246728

[23]    Shi K, Song W, Xu P, Liu R, Fang Z, Ji Y. Low-voltage ride-through control strategy for a virtual synchronous generator based on smooth switching. IEEE Access. 2017 Dec 18;6:2703-11. https://doi.org/10.1109/ACCESS.2017.2784846

[24]    Mukherjee S, Shamsi P, Ferdowsi M. Improved virtual inertia based control of a grid connected voltage source converter with fault ride-through ability. In 2016 North American Power Symposium (NAPS) 2016 Sep 18 (pp. 1-5). IEEE. https://doi.org/10.1109/NAPS.2016.7747871

[25]    Ma S, Geng H, Liu L, Yang G, Pal BC. Grid-synchronization stability improvement of large scale wind farm during severe grid fault. IEEE Transactions on Power Systems. 2017 May 2;33(1):216-26. https://doi.org/10.1109/TPWRS.2017.2700050

[26]    Huang L, Zhang L, Xin H, Wang Z, Gan D. Current limiting leads to virtual power angle synchronous instability of droop-controlled converters. In 2016 IEEE Power and Energy Society General Meeting (PESGM) 2016 Jul 17 (pp. 1-5). IEEE. https://doi.org/10.1109/PESGM.2016.7741667

[27] Zhang H, Xiong L, Gao Z, Yu S, Zhang X. Power matching based current limitation method for grid forming converter under large disturbances. International Journal of Electrical Power & Energy Systems. 2024 Jun 1;157:109841. https://doi.org/10.1016/j.ijepes.2024.109841

[28] Afshari E, Moradi GR, Rahimi R, Farhangi B, Yang Y, Blaabjerg F, Farhangi S. Control strategy for three-phase grid-connected PV inverters enabling current limitation under unbalanced faults. IEEE Transactions on Industrial Electronics. 2017 Jul 31;64(11):8908-18. https://doi.org/10.1109/TIE.2017.2733481